\documentclass[12pt,preprint]{aastex}

\begin{document}
\title{HST Imaging in the Chandra Deep Field South: I. Multiple AGN Populations}

\author{Ethan J. Schreier, Anton M. Koekemoer, Norman A. Grogin}
\affil{Space Telescope Science Institute, 3700 San Martin Drive, Baltimore,
MD 21218, USA}
\author{R. Giacconi, R, Gilli, L. Kewley, C. Norman}
\affil{Department of Physics and Astronomy, Johns Hopkins University,
Baltimore, MD 21218, USA}
\author{G. Hasinger}
\affil{Astrophysikalisches Institut, An der Sternwarte 16, Potsdam 14482
Germany}
\author{P. Rosati}
\affil{European Southern Observatory, Karl-Schwarzschild-Strasse 2,
Garching, D-85748, Germany}
\author{A. Marconi, M. Salvati}
\affil{Osservatorio Astrofisico di Arcetri, Largo E. Fermi 5, 50125
Firenze, Italy}
\author{P. Tozzi}
\affil {Osservatorio Astronomico di Trieste, Via G. B. Tiepolo 11,
34131 Trieste Italy}

\begin{abstract}
We present preliminary results from imaging three {\sl HST}/WFPC2
fields in $V$ and $I$ within the Chandra Deep Field South (CDFS). {\sl HST}'s
sensitivity and resolution are sufficient to reveal optical counterparts
for 24 of the 26 CDFS X-ray sources detected in the 300~ksec X-ray catalog
and to determine the morphologies of most of these. We find that the X-ray
sources comprise two apparently distinct populations of optical
candidates: one optically faint ($I\gtrsim24$) with $V\!-\!I$ colors
consistent with the $I>24$ field population; the other significantly
brighter ($I\lesssim22$) with colors redder than the $I<22$ field
population. More than 2/3 of the X-ray source counterparts are resolved
galaxies. The brighter sources are mostly AGN, based on their high X-ray
luminosity. The optically resolved sources in the brighter population have
a very narrow range of $V\!-\!I$ color and appear to be a mix of both late
and early type morphologies at low to moderate redshift.  We show that the
second population, with fainter optical counterparts, can be explained as
higher redshift Type~2 AGN.
\end{abstract}

\received{15 March, 2001}
\revised{10 May, 2001}
\accepted{15 May, 2001}

\section{Introduction}

Galaxy formation and evolution are fundamental topics in astrophysics
today. {\sl HST}'s high resolution has been particularly important in
opening up these fields and framing the questions: How did large scale
structure initially form? What is the role of hierarchical mergers?
What is the role of the massive central black holes (BHs) which now
appear to be found in most nearby galaxies (Magorrian et al.~1998)?

Just as high-resolution optical and near-IR observations are essential
to the study of galaxy formation and evolution, X-ray astronomy has
been key to the study of active galactic nuclei (AGN). We now commonly
explain AGN via accretion onto massive black holes, with the
difference between Type~1 and Type~2 AGN explained via an obscuring
torus in the standard unified model (e.g., Urry \& Padovani 1995).  We
further explain the cosmic X-ray background (XRB) via integrated
emission from AGN.  During the last few years, deep soft (ROSAT) and
hard (ASCA, BeppoSAX) X-ray surveys have provided essential
information on the evolution of AGN and on the origin of the XRB.
About 80\% of the $0.5-2$~keV background was resolved into discrete
sources by ROSAT (Hasinger et al.~1998), and most of these sources
were identified as AGN (Schmidt et al.~1998). These AGN are typically
broad line QSOs or Type~1 Seyferts, to be expected given ROSAT's
sensitivity to soft X-rays and thus unobscured AGN. The deep ROSAT
surveys also sampled the low end of the luminosity function of AGN at
high-$z$, discovering strong {\it density} evolution (Hasinger et
al.~1999, Miyaji et al.~2000). This is thought to be a consequence of
the larger rate of interactions and distorted/irregular morphologies
at high-$z$ (Lilly et al.~1998, Abraham et al.~1999).  Chandra is now
extending these results to lower fluxes and to harder X-rays
(cf.~Giacconi et al.~2001a, Hornschemeier et al.~2001, Mushotzky et 
al.~2000, Giacconi et al.~2001b, Tozzi et al.~2001).

Nonetheless, the physical nature of the relation between AGN evolution
and galaxy evolution is unknown.  Recent evidence for the existence of
this relationship comes from the strong correlation observed between
BH mass and host galaxy bulge velocity dispersion (Gebhardt et
al.~2000, Ferrarese \& Merritt 2000).  But the very basic question of
whether nuclear activity drives galaxy evolution or vice versa is
still open.

Answering these questions requires observing obscured and unobscured
AGN {\it and} their hosts, as a function of redshift.  The combination
of {\sl HST} and Chandra provides the potential to first detect AGN
via their X-ray emission, and then to study their optical host
galaxies with {\sl HST}.  This permits studying the effect of
environment on AGN and their evolution, and perhaps the effect of the
AGN on their environments.  The immediate questions are: 1) are tidal
interactions and/or distorted morphologies more prevalent in AGN hosts
than in field galaxies?  2) are host galaxy characteristics correlated
with AGN characteristics like flux and hardness (i.e., obscured
vs.~non-obscured), and do these correlations change with $z$?  3) can
we confirm the unified model at high $z$?  4) can we measure AGN
luminosity vs.~expected Eddington luminosity as a function of $z$ (by
estimating BH masses via follow-up velocity dispersion measurements of
host bulges)?

In this paper we show the first results from {\sl HST} imaging in the
Chandra Deep Field South (hereafter CDFS; Giacconi et al.~2001a). These
optical observations represent a pilot project to establish the
feasibility of studying faint X-ray source host galaxy morphologies.
Although the X-ray data have not yet been fully analyzed, and only
preliminary optical spectra and redshifts are available, we report several
new results from the preliminary analyses.  We confirm that we can
readily resolve structure in optical candidates for the CDFS sources.

In Section 2 we describe the observations and data reduction and in
Section 3, the optical source extraction.  In Section 4, we describe
the characteristics of the optical counterparts of the X-ray sources,
and in Section 5 we discuss the results. Throughout this paper we
assume a cosmology with $\Omega_\Lambda = 0.7$, $\Omega_{\rm M} =
0.3$, $H_0 = 70$~km~s${}^{-1}$~Mpc${}^{-1}$.

\section{{\sl HST}/WFPC2 Observations of the Chandra Deep Field South}

The optical observations presented here comprise three {\sl HST} Wide Field
Planetary Camera~2 (WFPC2) fields, located in the central regions of the
$\sim\!0.1$ deg$^2$ Chandra Deep Field South (CDFS). The CDFS is centered at RA
03 32 28.0, Dec $-$27 48 30 (J2000). It was selected for this very deep X-ray
survey because of its high galactic latitude, lack of bright stars, low 
$N({\rm H})$, and accessibility to VLT and Gemini.  The initial published X-ray
results (Giacconi et al.~2001a) were based on a total exposure of 126~ksec.
Our current analysis of the {\sl HST} observations uses the X-ray catalog and
results based on 300~ksec of exposure (Tozzi et al.~2001).  A total of about
940~ksec of Chandra data have by now been taken on the CDFS (cf.~Giacconi et
al 2001b). A forthcoming paper will extend our analysis to all X-ray sources
seen in the full 940~ksec CDFS exposure and will also present more detailed
quantitative morphology analyses for the optical counterparts.

Our {\sl HST}/WFPC2 exposures were taken with the F606W (hereafter $V$) and
F814W (hereafter $I$) filters (Table~1).  Five orbits were allocated to each
field: 2 orbits in $V$ and 3 in $I$.  The data were taken in July 2000 using
the ``dither'' mode, which involves offsetting the telescope by integral plus
1/2-pixel increments to both improve the sampling of the {\sl HST} PSF and
ameliorate the effect of bad pixels.  The data underwent standard WFPC2
pipeline calibration, after which the individual dithered images of each field
were combined into a single 0\farcs05/pixel image for each filter using the
{\sl drizzle} software (Fruchter \& Hook, 2001) in the IRAF/STSDAS {\sl dither}
package. The data were calibrated in the VEGAMAG photometric system, using the
current best values for the zero points (cf.~{\sl HST}/WFPC2 Instrument
Handbook, V.5, 2000).

Figure~1 shows color images of each of the three WFPC2 fields of view, with
blue representing F606W $V$, red representing F814W $I$, and green a mean image
of the two.  The three WFPC2 fields include a total of 26 X-ray sources listed
in the 300~ksec Chandra catalog.  Figure~2 shows the three WFPC2 fields of view
overlaid on a map of the 300~ksec X-ray sources, coded for intensity, as
detected by Tozzi et al.~(2001) in their $0.5-7$~keV detection band. Since we
chose HST pointings to optimize the number of X-ray sources observed, the
density of sources in our fields is higher than would be expected for random
WFPC2 fields.  Note that Tozzi et al.~find 197 sources in their 0.104~deg${}^2$
field, with clustering on scales up to 100\arcsec.  Our 26 sources in a total
area of $\approx\!17$ arcmin${}^2$ represents $\sim\!3$ times the average
source density.  Figure~3 is a collage of $20\arcsec\times20\arcsec$ {\sl HST}
$V\!+\!I$ image sections, centered on each Chandra source in our fields and
overlaid with the $0.5-7$~keV X-ray contours from the 300~ksec data (Tozzi et
al.~2001). We show, for reference, the expected {\sl CXO} point source response
size (FWHM) at each source location.

\section {Optical Source Extraction and Photometry}

We first created a detection image for each WFPC2 field by adding the $V$
and $I$ images.  We then extracted sources with the SExtractor program
(version 2.1.6; Bertin \& Arnouts 1996).  In light of the correlated
noise and the $\approx\!3$-pixel FWHM in the 0\farcs05/pixel drizzled
images, we adopted conservative SExtractor detection criteria: 10
contiguous pixels at $\geq\!1.5$ times the RMS ($\approx\!2.3$ counts
in the WFC regions and $\approx\!5.8$ counts in the PC regions).  For
a source with equivalent counts in each filter ($V\!-\!I=1.2$), this
corresponds to a WFC detection limit of $I \leq 28.2$ and $\mu_I \leq
24.2$ mag arcsec${}^{-2}$.  The corresponding limits for the PC are $I
\leq 27.2$ and $\mu_I \leq 23.2$ mag arcsec${}^{-2}$.

We ran SExtractor twice for each field using a single detection image as
described above, but performing the photometry separately on the $V$ and $I$
images.  This ensures that the source pixels are identically matched between
the two filters.  Our resulting optical source catalog contains 3834 objects
among the three fields.  We plot their $I$ magnitudes and $V\!-\!I$ colors
(based on the SExtractor MAG\_BEST parameter) as the small dots in
Figure~\ref{fig:col_mag}.  These magnitudes do not include correction for the
WFPC2 charge-transfer inefficiency (Whitmore, Heyer, \& Casertano 1999 and
references therein), which we expect to be $<\!0.03$~mag given the appreciable
background in our exposures.

We assessed the catalog completeness as a function of magnitude by repeating
the source extraction procedure after adding simulated point sources to the
image using the IRAF task {\sl mkobjects} in the {\sl ARTDATA} package. To
avoid overcrowding the images, we only added 150 simulated objects per WFC
region and 50 objects per PC region, iterating to improve the statistics. We
estimate that our catalog is complete to $I \approx 26.4$, where our
point-source recovery rate has fallen to $\approx\!90$\% for objects with the
median color $V\!-\!I\approx1.2$.  We plot this completeness limit as a
function of $I$ magnitude and $V\!-\!I$ color on Figure~\ref{fig:col_mag}
(dotted curve).  The simulated point-source recovery also gives us an
independent estimate of the SExtractor photometric errors. In the figure we
show $1\sigma$ photometric error bars for objects with the median colors at
$I$ = 23, 24, 25, and 26.  For brighter simulated sources, both $I$ and
$V\!-\!I$ are recovered to better than 0.03~mag, the level at which we expect
unmodeled systematic errors (e.g., charge-transfer inefficiency) to contribute.

\section {Optical Counterparts of X-ray Sources}

Each of our fields has $7-10$ objects clearly associated with CDFS 300~ksec
cataloged sources.  We registered each WFPC2 mosaic independently to the 
{\sl CXO} frame by subtracting the error-weighted median offset to these
{\sl CXO} sources.  We list the {\sl CXO} coordinate IDs of these counterparts
in Table~2, along with the coordinate offsets from their associated {\sl HST}
sources.  The mean deviation of the 24 registered {\sl HST} coordinates about
the {\sl CXO} source positions is $0\farcs4$, with a maximum offset of
$1\farcs2$.  This is broadly consistent with the {\sl CXO} positional
uncertainties, which range from 0\farcs1 to 0\farcs5 (Tozzi et al.~2001).  A
detailed analysis of the CDFS X-ray source positions with respect to the
optical galaxy positions --- in particular whether they all lie at the centers
of the galaxies --- will be valuable in helping determine the nature of the
X-ray emission (cf.~Brandt et al.~2001).  This will await the analysis of the
940~ksec data, where the expected greater number of sources will provide better
registration between the X-ray and optical fields, and where the better counting
statistics will improve the individual X-ray source positions.

There are only two CDFS 300~ksec sources (IDs J033204.5$-$274644 and
J033205.4$-$274644) within our three WFPC2 fields more than 2\arcsec\ away from
the nearest source in our catalog, 5~times the mean offset for the rest of the
sample. Both sources are on the PC in field CDFS2 and have $3\sigma$ upper
limits on $(V,I)$ of (26.8, 25.5) based on 0\farcs5-radius apertures centered
on the {\sl CXO} coordinates.  If these sources had fallen on one of the WFCs
rather than the PC, the detection upper limits would have been $\approx\!1$~mag
fainter.

For all 26 {\sl CXO} sources in our fields, we list in Table~2 the $V$ and $I$
magnitudes (or $3\sigma$ upper limits) from this study, as well as the derived
{\sl CXO} fluxes in the bands $0.5-2$~keV (column $F_{\rm XS}$) and $2-10$~keV
(column $F_{\rm XH}$), and the X-ray hardness ratio $HR$ from Tozzi et
al.~(2001). We also tabulate the hardness ratio $HR\equiv(H-S)/(H+S)$, where
$H$ and $S$ are the net counts in the hard ($2-7$~keV) and soft ($0.5-2$~keV)
band, respectively.  For comparison, the hardness ratio of the X-ray
background, with photon index $\Gamma\approx1.4$, is $HR\approx-0.38$ (Tozzi et
al.~2001).  In Table~2 we also include preliminary estimates for the redshift
and spectral classification for many of the optical counterparts based on
recent VLT/FORS spectroscopy by Hasinger et al.~(2001).  Because emission-line
ratio diagnostics were not yet available to discriminate between Type~2 AGN and
starburst spectra, we conservatively label all such candidates as narrow
emission-line galaxies (NELGs).

We find that the SExtractor star/galaxy separation is reliable to
$V\lesssim25.4$, substantially fainter than most {\sl CXO} counterparts.
Values of the $V$-band SExtractor STELLARITY parameter $\eta_{V}$ span the
range $0-1$, with clearly resolved sources having $\eta_{V}\lesssim0.2$ and
clearly unresolved sources having $\eta_{V}\gtrsim0.8$.  Fifteen of the 24
optical counterparts to the X-ray sources are clearly resolved
($\eta_{V}<0.2$).  Most of these resolved sources are sufficiently extended to
permit detailed examination and classification as elliptical, spiral, or
irregular galaxies --- we note their optical morphology in Table~2. Of the
remainder, seven are clearly unresolved, and two are too faint for reliable
star/galaxy separation (listed as ``indeterminate'' morphology in Table~2).

For the clearly extended sources, we also fit surface brightness profiles
with the IRAF/STSDAS isophotal analysis package, ISOPHOTE.  The package's
contour fitting task {\sl ellipse} takes an initial guess for an isophotal
ellipse, then steps logarithmically in major axis. At each step it finds the
optimal isophotal ellipse center, ellipticity, and positional angle. Prior to
the ellipse fitting, we use the task {\sl imedit} to mask foreground stars,
neighboring galaxies, etc., near the galaxies of interest, and ignore masked
pixels in the fitting. The surface brightness profiles corroborate the
qualitative morphology assessment obtained by visual inspection.  The
quantitative results from the surface photometry will be presented in the
forthcoming paper.

In Figure~\ref{fig:col_mag}, we flag the optical counterparts of the X-ray
sources among the full-field color-magnitude diagram with symbols according to
their optical morphology: circles represent resolved early type galaxies (types
S0 and earlier); squares are resolved later type galaxies (types Sa and later);
stars represent clearly unresolved sources; and triangles are objects of
indeterminate morphology.

\section{Discussion}

There is a readily apparent dichotomy in the color magnitude distribution
of the optical counterparts of the X-ray sources: a brighter group with
$I \sim 17.8-22.7$, and a fainter group with $I \sim 24.2-25.3$. The
fainter group (containing only five sources) is consistent in its color
distribution with the field galaxy population. The other group is both
significantly brighter and significantly redder, on average, than the
population of field sources.

Further, those optically bright X-ray sources that are extended in our {\sl HST}
images (those indicated by squares and circles in Figure~\ref{fig:col_mag})
appear redder than the rest of the bright sources. This subset of bright
extended sources occupies a narrow range of color, $V\!-\!I \sim 1.7-2.2$,
while spanning a magnitude range $I \sim 19-22.7$.  There is also a suggestion
that the Chandra-detected late type galaxies are optically fainter than the
corresponding early-type galaxies.

The fact that the X-ray sources with resolved optical candidates appear to
occupy two different regions of color-magnitude space strongly suggests
that we are seeing two (or more) different populations of sources. We note
that the faint limit ($I\sim 23$) of the brighter population is well above
our completeness limit, and that the distributions of these sources in both
magnitude and color are significantly different from the field population
distributions. In Figure~\ref{fig:col_hists}, we plot histograms of the number
of sources vs. color for the X-ray source counterparts (solid) and the field
sources (dotted). The top set of histograms includes sources at all magnitudes,
the bottom is limited to sources with $I<23$. The redder color of the X-ray
emitting galaxies is obvious, even when we limit the field sample to the
brighter sources.  A Kolmogorov-Smirnov (K-S) test rules out a common
underlying color distribution at the $4\sigma$ level.

The color-magnitude diagram also shows a far greater incidence of X-ray
counterparts for the optically brighter sources than for the fainter sources
--- 16 {\sl CXO} detections among the 195 {\sl HST} sources with $I \leq 22$,
in contrast to 5 detections among the 2060 sources with $26 \leq I \leq 23$.
We suggest that the brighter population is spatially relatively sparse, since
the sources are observed well above the optical detection limit. Furthermore,
the X-ray luminosities estimated from the available preliminary redshifts
suggest that these are predominantly AGN: 11 of 16 optically bright sources
have X-ray luminosities $L(0.5-10\,{\rm keV}) > 10^{43}$~erg~s$^{-1}$, and 5
have luminosities in the range $10^{42} - 10^{43}$~erg~s$^{-1}$, likely Type~2
AGN and/or starburst galaxies. The fact that the AGN are associated, on
average, with redder galaxies could suggest the presence of dust associated
with mergers in AGN hosts.

The X-ray emitting galaxies in the optically fainter group represent a
small fraction of the field galaxies, but their colors, on the other hand,
are near the average for the field population.  Their optical fluxes are
not much above our detection limits, and we suggest that these could well
represent the high X-ray intensity tip of a much larger population of
sources.

In Figure~\ref{fig:col_mag_model}, we overlay predicted templates for
elliptical, spiral, and irregular/starburst galaxies on the same
color-magnitude diagram, motivated by similar comparisons carried out
previously, for example in the HDF-N (Williams et al.~1996). The templates are
taken from Coleman et al. (1980), with fluxes computed through the {\sl  HST}
filter bandpasses after applying a range of redshifts, and assuming $\sim\!L_*$
luminosities for each galaxy class. We also plot a Type~1 AGN track (for a QSO
of $M_B = -24$), assuming the fluxes are dominated by the flat-spectrum nuclear
component. For the X-ray counterparts, we use the same symbols as in
Figure~\ref{fig:col_mag} to distinguish between ellipticals, spirals,
unresolved and indeterminate optical classifications. We further code for the
intensity and hardness ratio for the X-ray sources, with size being
proportional to X-ray flux, and darker shading indicating harder X-ray spectra.
We see from the tracks that the bright, extended population of objects in the
narrow color range $V\!-\!I \sim 1.9$ are consistent with early-type galaxies
and spirals at low to moderate redshifts ($z \sim 0.5 - 1.2$).
Our optical data confirm that we see both early-type and spiral morphologies
for these objects, and those preliminary redshifts which are available
(Hasinger et al.~2001) are also consistent with the expected range (cf.
Table~2). The range of apparent magnitudes ($I \sim 19-23$) is consistent with
these objects being $\sim\!L_*$ galaxies at these fairly moderate redshifts.
We note that the spirals in this sample are generally fainter than the
ellipticals, in agreement with the predictions from the redshifted tracks.
Finally, we find that the unresolved sources generally fall between the
elliptical/spiral template tracks and those of the pure Type~1 AGN
spectrum, suggesting that these objects are Type~1 AGN with smaller, varying
admixtures of host galaxy contributions to the total flux. These point-like
sources are also bluer, for a given magnitude, than the resolved sources.
The colors of these objects are similar to starbursts, however their X-ray
luminosities are higher than those expected for typical starbursts, thus we
conclude that they are more likely to be AGN.

We have already noted that the color distribution of the fainter group is
consistent with that of the field galaxy population. The predicted galaxy
tracks suggest that the field population at these magnitudes consists of
irregulars and more distant spirals and starbursts.  We may also be
starting to see the faint blue population of galaxies that has been
previously discussed in the context of the HDF-N (e.g.,~Mobasher et
al.~1996, Phillips et al.~1997). The X-ray sources in this optically faint
population have high $F_{\rm X}/F_{\rm opt}$ and are likely to be distant
AGN, as discussed below.

To further explore the nature of the two populations, we have simulated the
expected number counts and color-magnitude distribution for various classes
of objects. We have used published template SEDs, luminosity functions and
general evolution models for ellipticals, spirals, and starbursts, as well 
as Type~1 and Type~2 AGN. For ellipticals, spirals and starburst/Im galaxies
we used templates from Coleman et al.~(1980) at optical wavebands, combined
with multiwaveband data from Schmitt et al.~(1997) extending up to X-ray
energies. For Type~1 AGN we used composite spectra from Cristiani \& Vio (1990)
supplemented again by Schmitt et al.~(1997) for X-ray energies, while our
Type~2 AGN template was taken from Della Ceca et al.~(2000). For the
non-active galaxies we adopted the same luminosity function and evolution as
determined by Loveday et al.~(1992), while for the AGN we used the Type~1
luminosity function and evolution recently determined by Boyle et al.~(2000).
We applied the Type~1 LF to the Type~2 AGN by assuming that the ratio
of Type~1/2 remains constant with redshift, is determined by a torus with
opening angle 60$\arcdeg$, and that the emission from the torus is isotropic
at 60~$\mu$m (thus normalizing the SEDs at this frequency). These are all
fairly extreme oversimplifications, and we will present more detailed modelling
in Koekemoer (2001) in the context of the entire set of sources from
the 940~ksec catalog. For the moment, however, we note that the results are
not dramatically sensitive to details such as the precise ratio of Type~1/2
(and whether it varies with redshift). We simulated the expected fluxes in the
optical using the filter throughput curves for {\sl HST}/WFPC2, and similarly
the Chandra fluxes in the soft and hard bands were simulated using the
available sensitivity data for ACIS-I as a function of energy.

In Figure~\ref{fig:popns}, we present the histogram of the $I$ magnitude
distribution of the observed optical counterparts of the Chandra sources, along
with predicted curves (to be discussed in more detail our forthcoming paper)
for ellipticals, spirals, starbursts, and AGN, based on the above models, an
area of sky corresponding to our three WFPC2 fields (a total of
17.1~arcmin$^2$), and detection thresholds for the 300~ksec Chandra X-ray data
as described by Tozzi et al.~(2001). We also show the sum of the contributions
from each type.

It is clear from the predictions in Figure~\ref{fig:popns} that the majority of
the X-ray source optical counterparts are expected to be relatively bright, as
observed. The group includes ellipticals, spirals, starbursts, and Type~1 and
Type~2 AGN. More interesting, the optically faint population
($I\gtrsim 24 - 25$) of X-ray source counterparts is expected to consist only
of higher-redshift Type~2 AGN ($z \sim 2$). Furthermore, the models suggest
that these would be softer than their lower-redshift counterparts, by virtue of
the predicted higher redshifts causing more X-ray emission to be shifted into
the soft band. This prediction agrees with the observed X-ray hardness ratios:
the optically faint sources are not as hard in the X-rays as the optically
brighter ones. Moreover, the one object in this faint group for which we
have a preliminary redshift (J033208.3$-$274153) is a NELG with $z=2.45$,
leading to an X-ray luminosity estimate of $ \approx
3\times 10^{43}$~erg~s$^{-1}$, in the rest frame energy range of 
$1.7-35\,{\rm keV}$. This is substantially brighter than
expected for a starburst, and is likely a distant type 2 AGN.

In Figure~\ref{fig:X_opt_magmag}, we plot X-ray vs. optical flux of the CDFS
sources, for the hard and soft X-ray bands separately. The population of
optically brighter, resolved sources lies along a line of constant
$F_{\rm X}/F_{\rm opt}$ with a relatively low value compared with the other
CDFS sources. As stated earlier, the lack of objects at lower
$F_{\rm X}/F_{\rm opt}$, i.e. in the lower left corner of the plot, is not an
observational bias.  As we move to even lower values of $F_{\rm X}/F_{\rm opt}$,
we would eventually reach levels consistent with X-ray emission from normal
galaxies.

The unresolved sources tend to have higher $F_{\rm X}/F_{\rm opt}$, reflecting
the relative dominance of AGN emission over host galaxy contribution over
these mostly Type~1 AGN. Note also that the optical non-detections (lower
right) are consistent with the high $F_{\rm X}/F_{\rm opt}$ ratio of the
faint population.

For fainter X-ray flux detection limits, our models predict a substantial
increase in the number of optically faint counterparts, the majority of which
should be obscured AGN at reasonably high redshifts of $z \gtrsim 1.5-2.5$
(cf.~Figure~\ref{fig:popns}). While we can also expect to see additional
objects in the ``gap'' between the two populations, the ``double peak'' in the
source count histogram should persist -- the bright population of unobscured
AGN and starbursts, and the fainter population of distant, obscured AGN, with
the peak of the faint population being considerably broader than that of the
bright population.

Although few of the {\sl CXO}-detected galaxies show obvious signs of strong
disruption or disturbances (e.g., J033213.3$-$274241), most have nearby
companions ($\lesssim\!5$\arcsec) and at least a few reside within apparent
compact groups (e.g., J033208.1$-$274658, J033233.1$-$274548).  If galaxy
activity and X-ray emission were correlated with local environment, e.g., from
interaction-induced star formation and/or feeding of a central BH, we might
expect to see more near neighbors around the {\sl CXO}-detected galaxies than
typical for the field. We investigate this by determining the number of
companion galaxies ($\eta_{V}<0.2$), with $I\leq23$, within 8$\arcsec$ of
each $I\leq23$ galaxy in our source catalog.  For our adopted cosmology,
8$\arcsec$ corresponds to a projected distance of 50~kpc at $z\approx0.6$,
typical of the {\sl CXO}-detected galaxies at $I\leq23$ (cf.~Table~2).
Figure~\ref{fig:neighbors} shows the histogram of this near-neighbor
distribution for the 12 {\sl CXO}-detected $I\leq23$ galaxies (solid line) and
for the full sample of the 328 galaxies with $I<23$ (dotted line).  Although
the chi-square test gives a high probability (0.49) of the null hypothesis that
the two histograms are consistent with a single nearest-neighbor distribution function, the
analysis would greatly benefit from enlarging the {\sl CXO}-detected population
with wider optical coverage (and with the 940~ksec CDFS data).

We have verified that our selection of {\sl HST}-observed sources is, on
average, representative of the full catalog of 300~ksec CDFS sources in terms
of X-ray flux and hardness.  Figure~\ref{fig:xfluxhr} shows the hardness ratio
versus $0.5-10$~keV flux for the full 300~ksec catalog of X-ray sources,
overlaid with our standard symbols for the optical counterparts.  We see that
the optically resolved counterparts span a broad range of the X-ray parameter
space.  We also note that the unresolved counterparts appear, on average,
softer in the X-rays, reflecting a lack of obscuration.  As there are not yet
accurate redshifts for half of our optical counterparts, we cannot make
definitive statements based on intrinsic luminosities.

\section{Conclusion}

Summarizing our initial results, we find that with only two exceptions, all
X-ray sources detected by Chandra in the initial 300~ksec exposure that lie in
our WFPC2 FOVs have apparent optical counterparts within $\sim\!0.5$~arcsec, to
our limiting magnitude of $V\sim28.4$. The majority of the candidates are
resolved galaxies, and many are at low to moderate redshifts.

There appear to be two distinct populations of optical counterparts for the
X-ray sources: a brighter population, with $I \sim 17.8-22.7$; and a fainter
population, with $I \sim 24-25$.  The brighter population is both significantly
brighter and significantly redder on average than the population of field
sources. The fainter population (only five sources) is consistent in its color
distribution with the field galaxy population.  There is no significant
difference in X-ray hardness between the groups, although in each group, the
harder sources appear marginally redder.

The optically bright X-ray sources identified with resolved galaxies are redder
than the overall population of bright sources and lie in a narrow range of
color: $V\!-\!I \sim 1.7-2.2$, over a magnitude range of $I = 19-22.7$. These
values are consistent with $\sim\!L_*$ galaxies at moderate redshifts of 0.5 to
1.2, comparable to the range of preliminary redshifts already measured
(cf.~Table~2).  The high X-ray luminosities suggest that most of these harbor
AGN.  The redder colors of the {\sl CXO} optical counterparts further imply
that these AGN hosts are dusty, as might be expected if mergers were associated
with nuclear activity.  The spirals in this sample are also generally fainter
than the ellipticals, as expected from the comparison of their colors and
luminosities with predicted galaxy templates. The bright population of galaxies
lies well above our optical detection threshold, suggesting that these comprise
a relatively sparse, volume-limited and not flux-limited population.

The optically faint group of {\sl CXO} sources is consistent in color
distribution with the field galaxy population. These sources are likely to be
high redshift Type~2 AGN ($z \sim 2$).  They appear softer than the nearer
Type~2 AGN because their unobscured hard X-ray emission is substantially 
redshifted into the {\sl CXO} soft band.  We predict that at
still fainter X-ray flux limits, we will find substantially more of the faint
optical counterparts, but few additional members of the resolved brighter
population.

In conclusion, we note that these {\sl HST} observations cover only some
$7.5\%$ of the full-depth 300~ksec CDFS area (Tozzi et al.~2001).  More and
deeper observations should be undertaken.  Our current results suggest that
{\sl HST}/WFPC2 or ACS exposures of the CDFS at $\mu_I\sim24-25$ mag
arcsec$^{-2}$ can easily resolve the optical counterparts of many more of the
Chandra sources in our fainter population. This should represent a complete
sample of X-ray selected AGN host galaxies out to $z\sim 2$, at or beyond the
expected peak in AGN density evolution. {\sl HST}/NICMOS $K$-band observations
and the eventual {\sl SIRTF} Great Observatories Origins Deep Survey (GOODS) of
this field will provide valuable infrared diagnostics of the AGN vs.~starburst
contributions, in conjunction with additional ground-based optical and near-IR
spectroscopy.

These initial results already demonstrate the ability of {\sl HST}, in a
relatively modest amount of observing time, to readily detect the host
galaxies of distant, low-luminosity AGN and to reveal their morphology. We will
present more quantitative morphology results, detecting color gradients and
distinguishing unresolved AGN emission from the host galaxy emission, in a
forthcoming paper that will include the optical counterparts from the full
940~ksec CDFS exposure.

\acknowledgements
We gratefully acknowledge the award of {\sl HST} Director's Discretionary time
in support of this project. We also acknowledge support for this work which was
provided by NASA through GO grants GO-08809.01-A and GO-07267.01-A from the
Space Telescope Science Institute, which is operated by AURA, Inc., under NASA
Contract \hbox{NAS 5-26555}. We thank the referee for very useful suggestions.

\clearpage

\setcounter{figure}{0}
\begin{figure}
\caption{(a) {\sl HST} three-color images of each of the three WFPC2 fields
in the Chandra Deep Field South, made by assigning red to $I$, blue to $V$,
and green to the average of the two bands. North is to the top and East is to
the left. This figure is for the CDFS1 field.}
\end{figure}


\setcounter{figure}{0}
\begin{figure}
\caption{(b) As for Fig.~1a, but for the CDFS2 field.}
\end{figure}


\setcounter{figure}{0}
\begin{figure}
\caption{(c) As for Fig.~1a, but for the CDFS3 field.}
\end{figure}

\clearpage

\setcounter{figure}{1}
\begin{figure}
\plotone{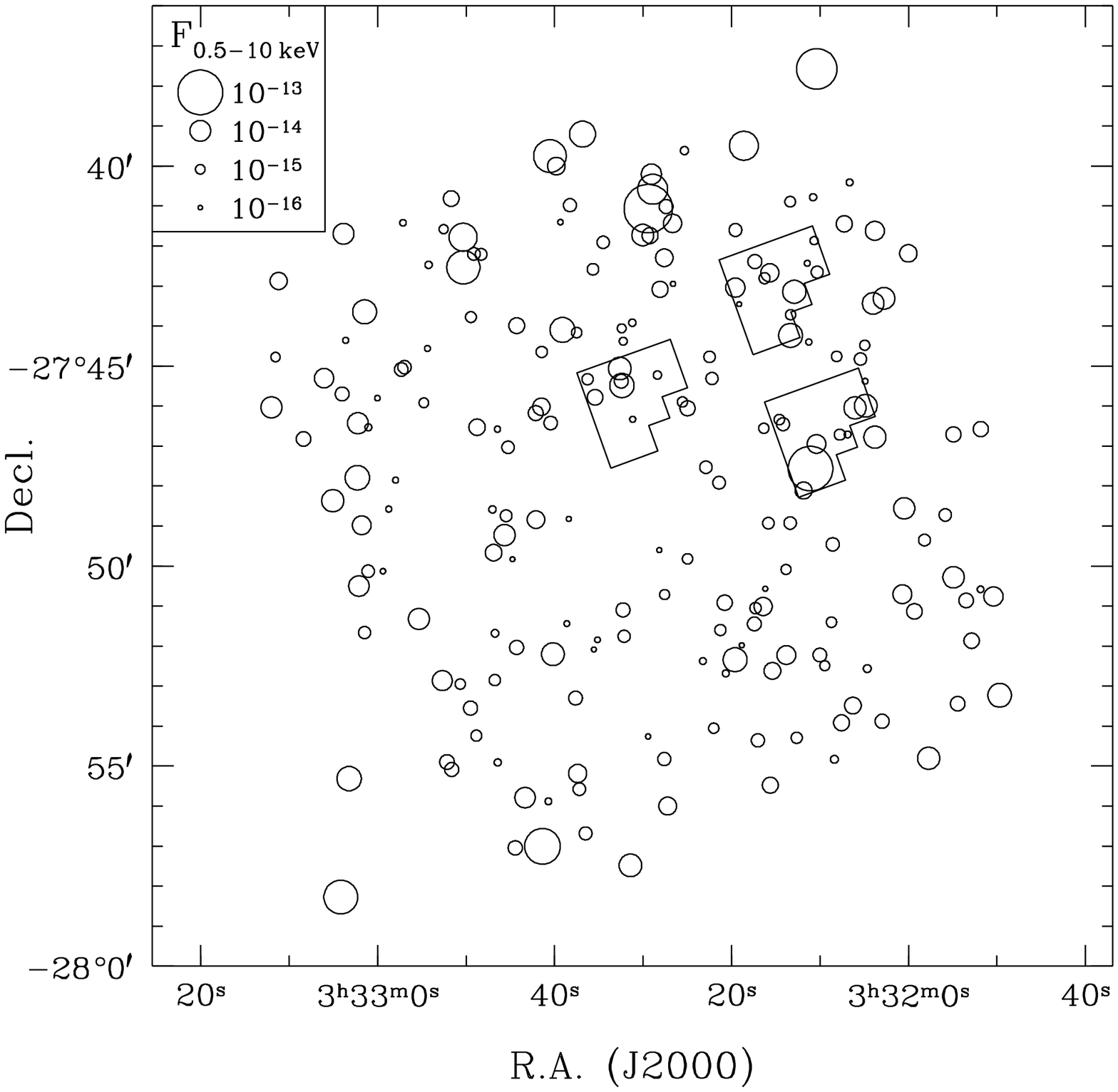}
\caption{
Map of all X-ray sources (open circles) detected in the 300 ksec {\sl CXO}
image of CDFS.  The three {\sl HST}/WFPC2 fields in this study are
outlined by the chevrons toward the upper right.  The {\sl CXO} source fluxes
(in erg s${}^{-1}$ cm${}^{-2}$ at $0.5-10$~keV) are denoted by symbol
size, as indicated by the legend at the upper left.
}
\end{figure}

\clearpage

\setcounter{figure}{2}
\begin{figure}
\caption{(a) {\sl HST} greyscale images of each of the X-ray sources in the
three CDFS WFPC2 fields, made by combining the $V$ and $I$ datasets. The X-ray
contours from the smoothed 300~ksec Chandra data are overlaid on each image
(with the first three contour levels corresponding to 1, 2, and 3~sigma,
and increasing thereafter by a factor of two per contour level). For each image
we display a scale bar indicating the FWHM of the ACIS-I PSF. North is to the
top and East is to the left.}
\end{figure}


\setcounter{figure}{2}
\begin{figure}
\caption{(b) As for Fig.~3a.}
\end{figure}


\setcounter{figure}{2}
\begin{figure}
\caption{(c) As for Fig.~3a.}
\end{figure}

\clearpage

\begin{figure}
\plotone{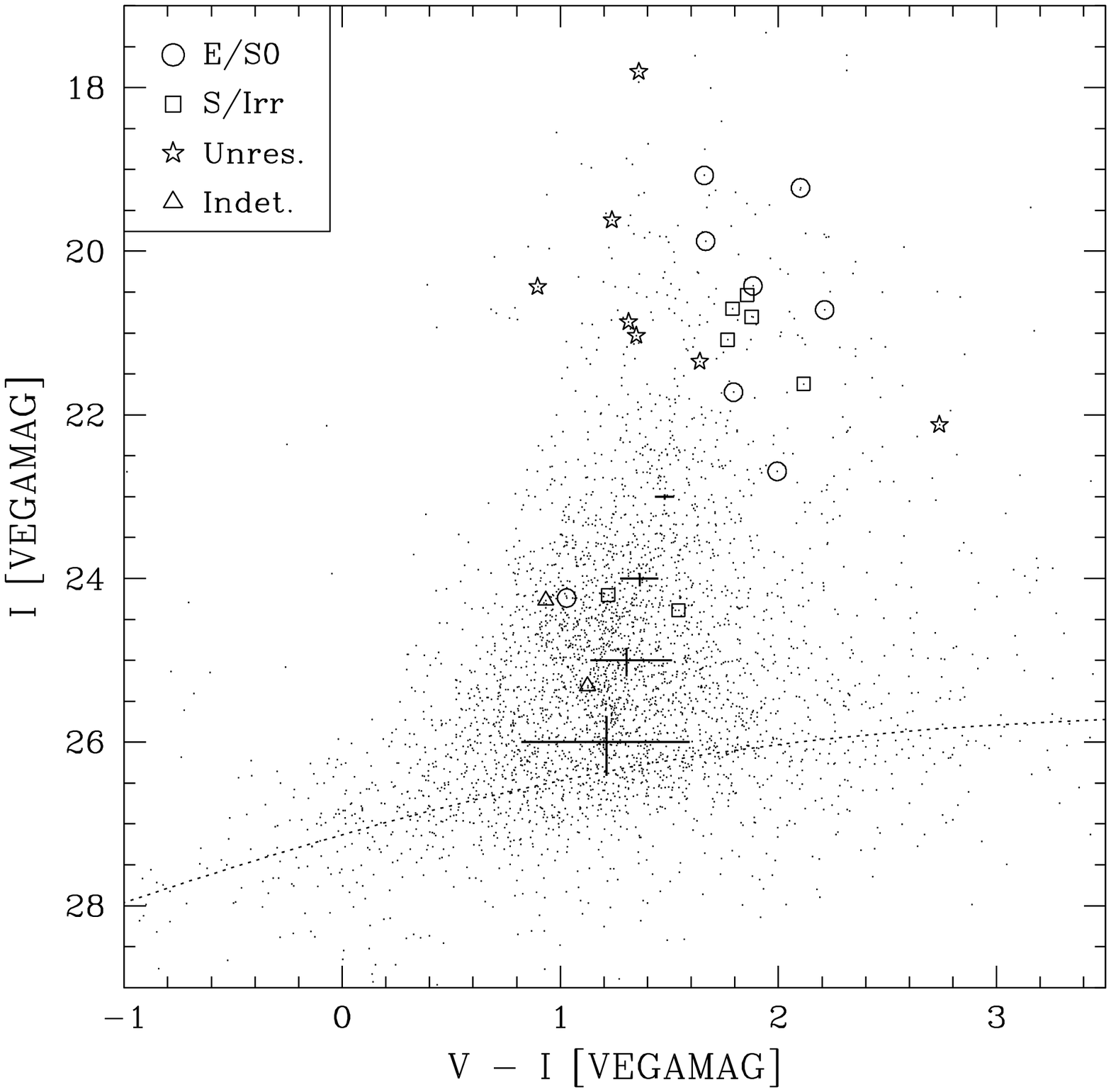}
\caption{\label{fig:col_mag}
Color-magnitude diagram of the 3681 sources (small dots) with both
$V$ and $I$ detections in the three WFPC2 fields of this study.
Sources with X-ray emission detected in the 300~ksec CDFS image are
flagged with larger symbols according to their optical morphology:
resolved galaxies of types S0 and earlier (circles); resolved galaxies
of types Sa and later (squares); unresolved sources (stars); and
indeterminate (triangles). We indicate the catalog completeness limit
(dotted line) as well as photometric error bars for sources with
$I$ = 23, 24, 25, 26 at the median color for the respective magnitude. }
\end{figure}


\begin{figure}
\plotone{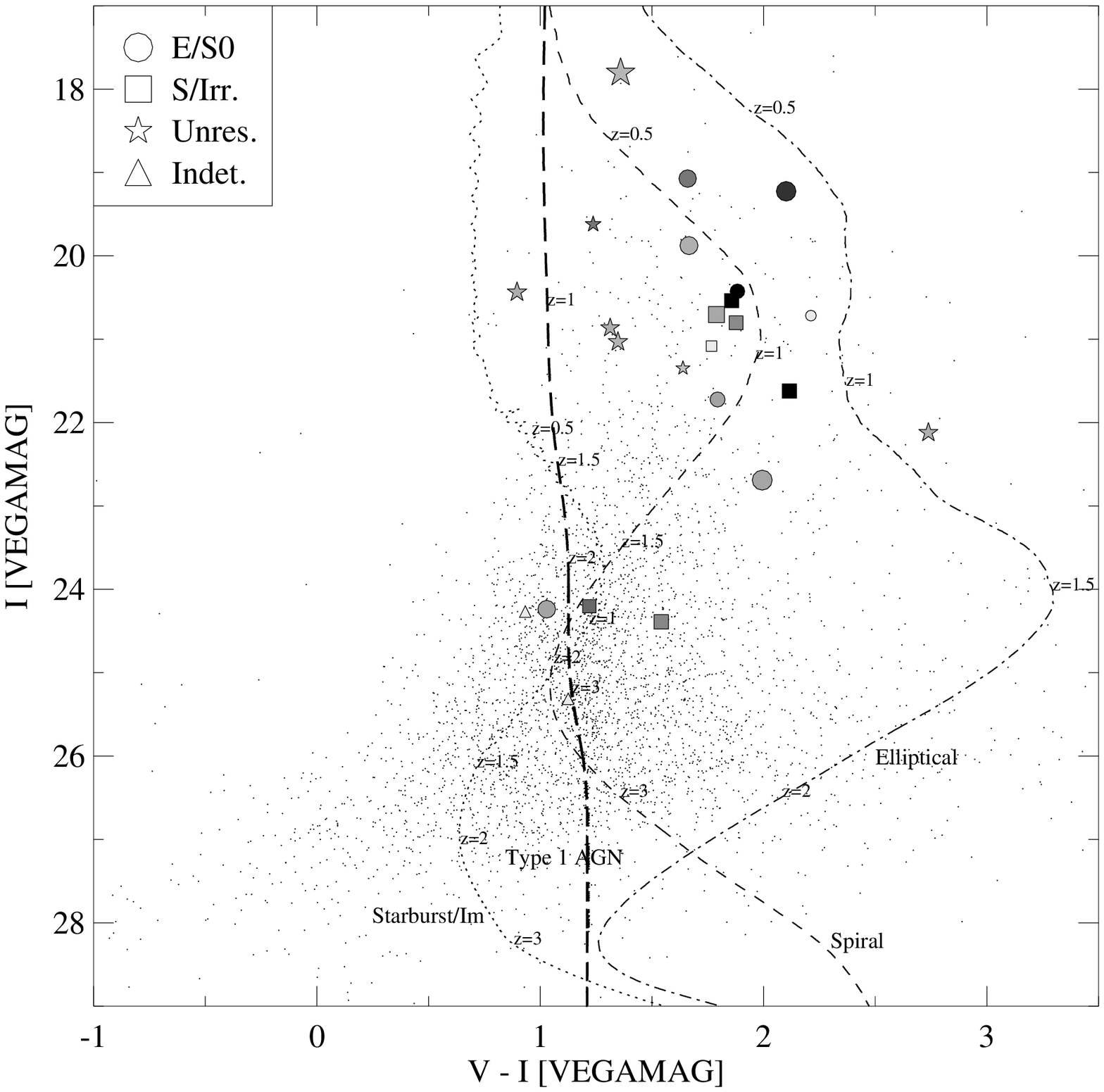}
\caption{\label{fig:col_mag_model}
Color-magnitude diagram for all sources, showing tracks for elliptical, spiral,
starburst/irregular and Type~1 AGN template galaxy SEDs as a function of
redshift; see text for further details. Symbols for X-ray detected sources are
the same as in Figure~\ref{fig:col_mag}, except that additional information is
contained in their size which now indicates total X-ray flux, while their
shading represents X-ray hardness ratio (ranging from white to black for
$HR=-1$ to $HR=+1$ respectively).}
\end{figure}

\begin{figure}
\epsscale{0.8}
\plotone{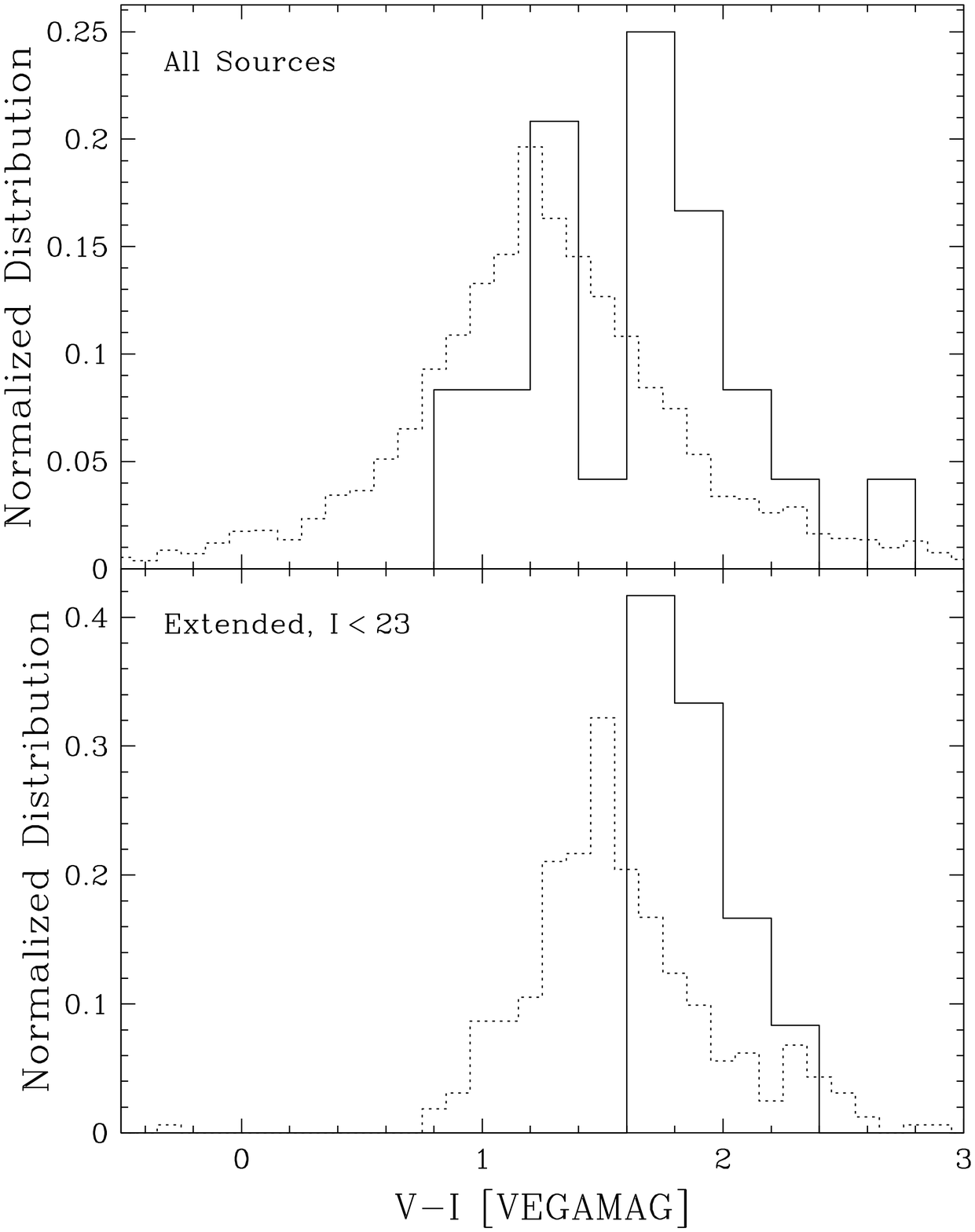}
\epsscale{1.0}
\caption{\label{fig:col_hists}
Color histograms of {\sl CXO} sources (solid; 0.2 mag bins) versus the entire
field (dotted, 0.1 mag bins). The upper plot shows all sources, while the lower
plot shows the subset of bright extended sources ($I<23$, $\eta_{V} < 0.2$).
The histograms are all normalized by bin size and sample size.}
\end{figure}


\begin{figure}
\plotone{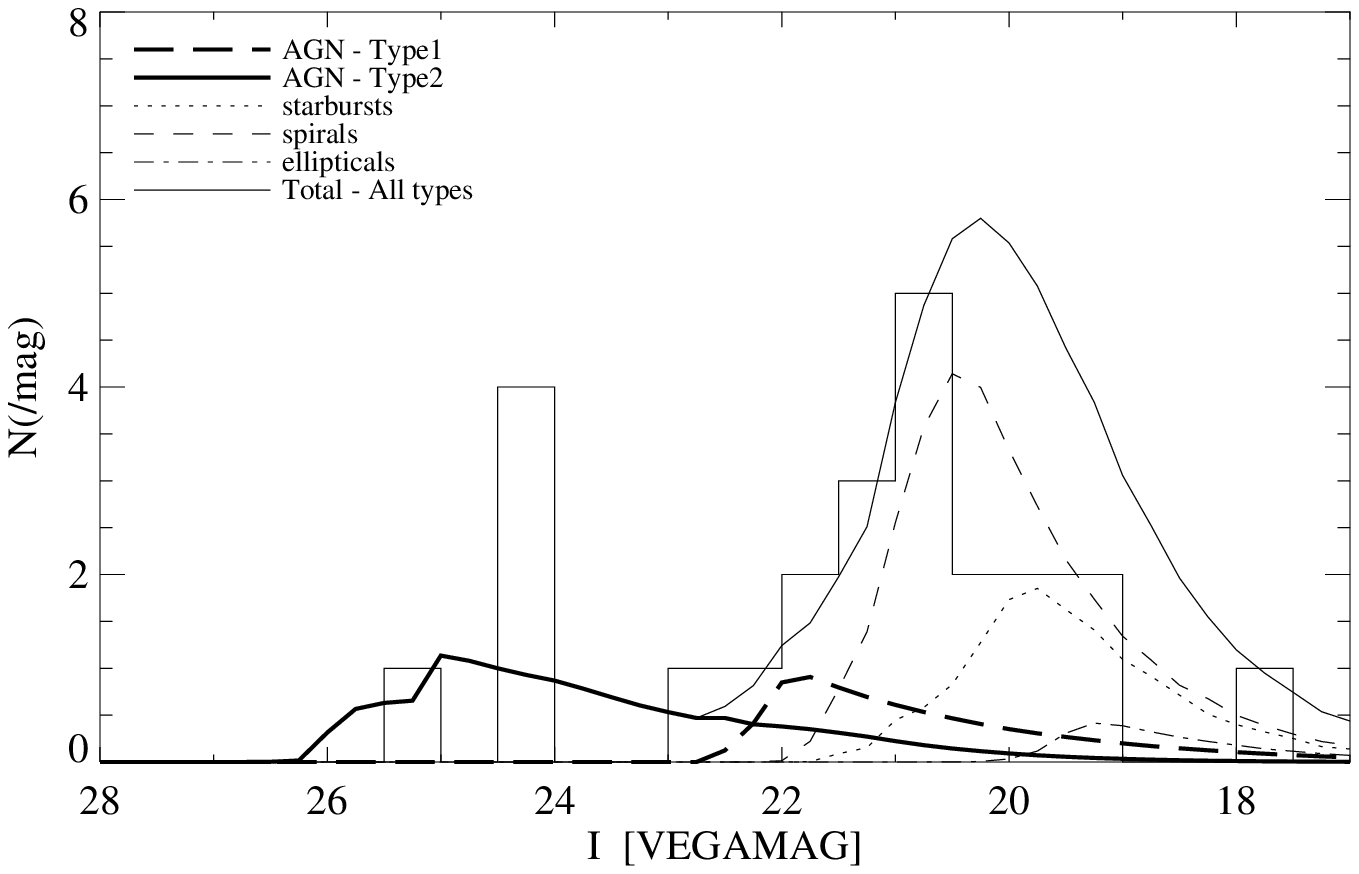}
\caption{\label{fig:popns}
Histogram of the $I$ magnitude of the sources with detected X-ray emission,
showing the contribution of various different classes of objects. Note in
particular that the spirals, ellipticals, and Type~1 AGN are all relatively
bright, while the Type~2 AGN are the only ones with a distribution that is
peaked fainter than $I\sim23$.}
\end{figure}


\begin{figure}
\epsscale{0.65}
\plotone{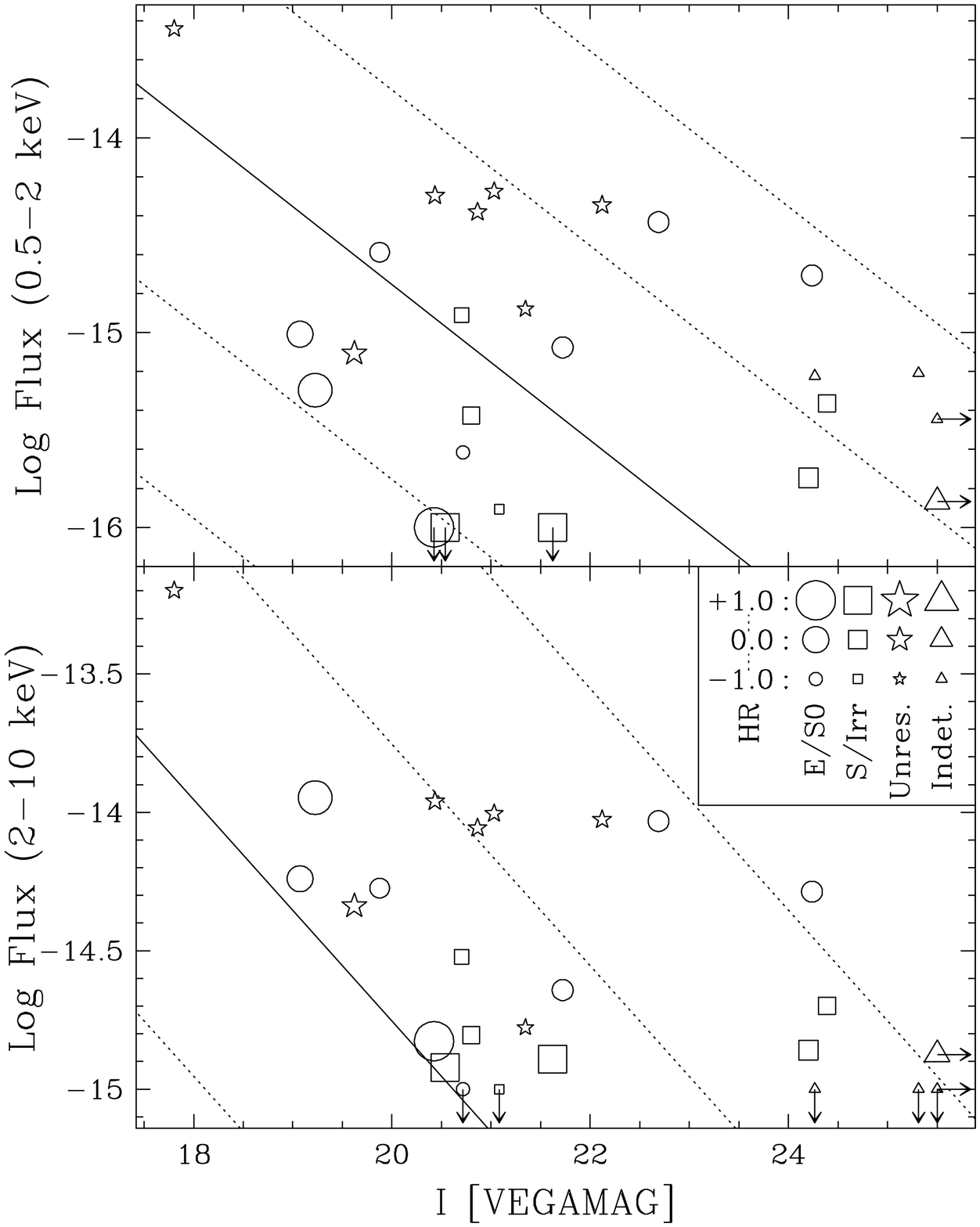}
\epsscale{1.0}
\caption{\label{fig:X_opt_magmag}
$I$ magnitude versus derived X-ray flux at $0.5-2$~keV (upper) and $2-10$~keV
(lower) for all CDFS 300~ksec sources.  The different symbols represent optical
morphology, as described in Figure~\ref{fig:col_mag}, and their sizes represent
X-ray hardness ratio (legend at right center; see text for definition).  In the
upper plot, sources detected only in hard X-rays ($2-10$~keV) are shown with
upper limits given the soft X-ray detection limit of
$10^{-16}$ erg~s$^{-1}$~cm$^{-2}$. In the lower plot, sources detected only in
soft X-rays ($0.5-2$~keV) are shown with upper limits given the hard X-ray
detection limit of $10^{-15}$~erg~s$^{-1}$~cm$^{-2}$. In both plots, {\sl CXO}
sources without {\sl HST} counterparts are shown with $3\sigma$ upper limits in
$I$.  The solid line across each plot represents $F_{\rm X}/F_{\rm I}=1$
assuming the conversion
$I$-band 0 mag = $1.762 \times 10^{-7}$~erg~s$^{-1}$~cm$^{-2}$.  The dotted
lines are spaced at 1~dex intervals in $F_{\rm X}/F_{\rm I}$.}
\end{figure}

\clearpage

\begin{figure}
\plotone{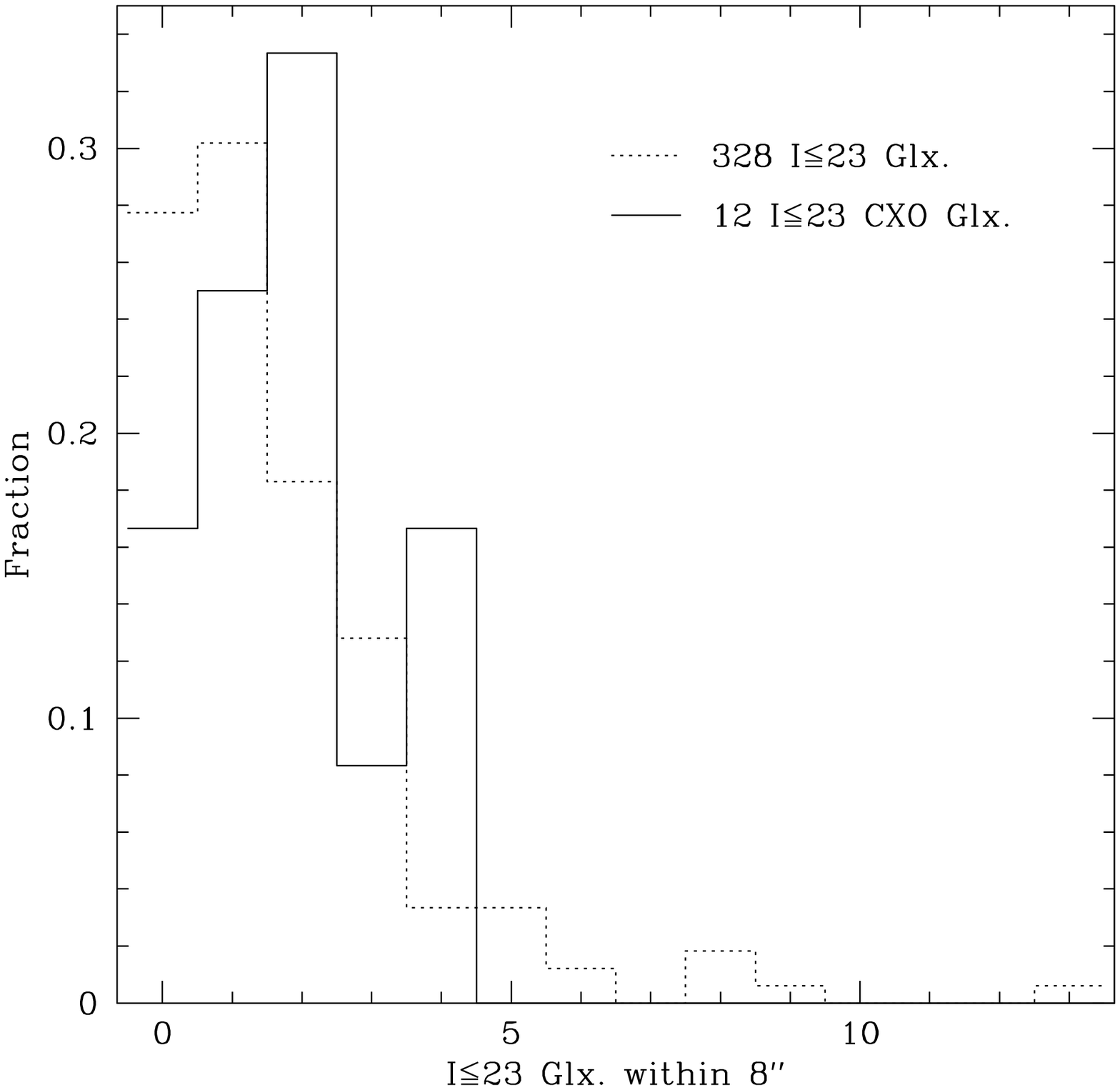}
\caption{\label{fig:neighbors}
Histogram of the number of nearby ($<\!8$\arcsec) $I<23$ galaxies appearing
around the 328 galaxies in our catalog with $I<23$ (dotted) and around the
subset of 12 $I\leq23$ galaxies detected by {\sl CXO} (solid).  The histograms
are normalized by the sample sizes.}
\end{figure}

\clearpage

\begin{figure} 
\plotone{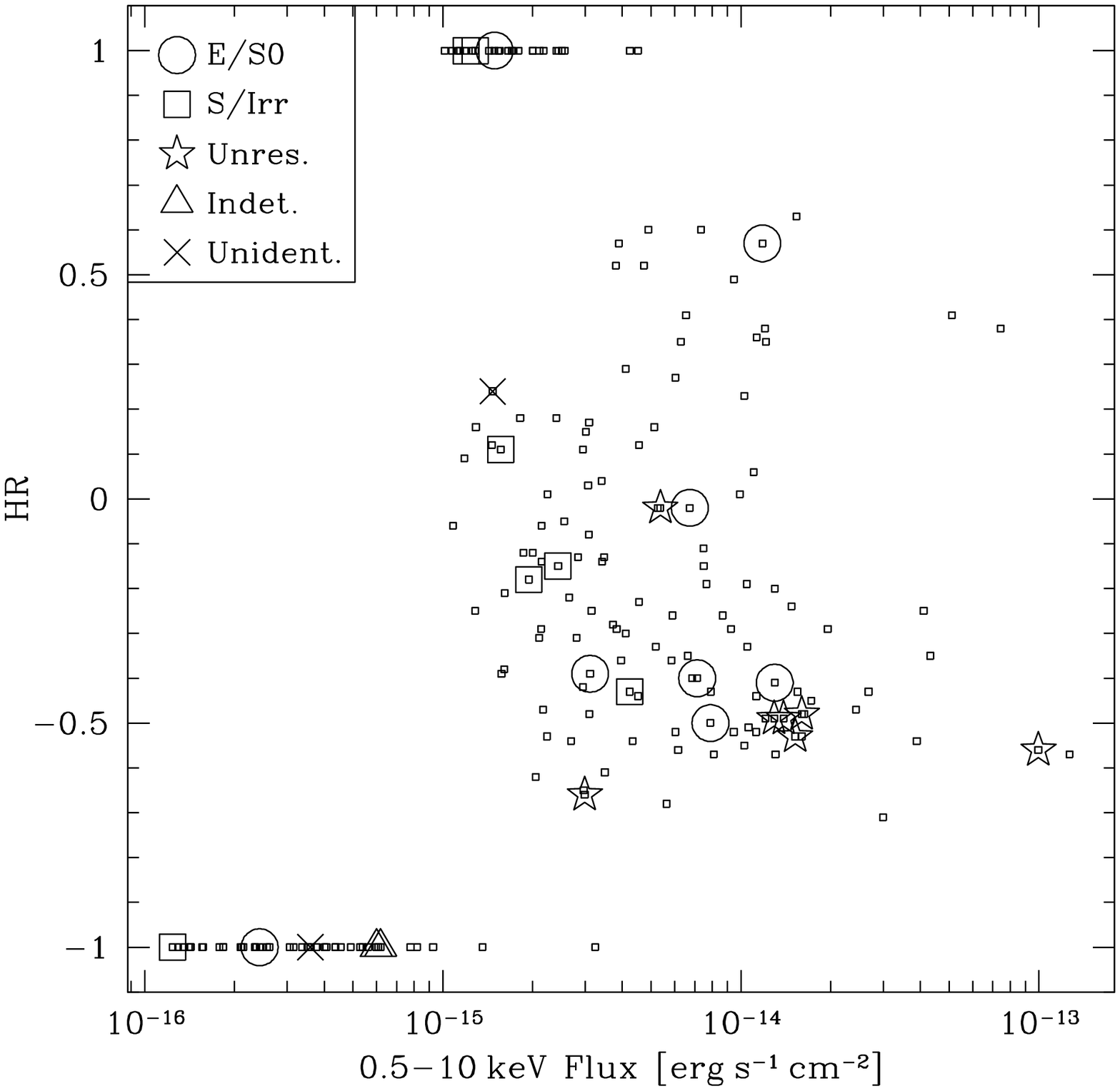} 
\caption{\label{fig:xfluxhr}
Hardness ratio versus $0.5-10$~keV flux for all {\sl CXO} sources detected in
the 300 ksec CDFS image (small squares).  The larger symbols denote the
morphologies of the optical counterparts within the three 
{\sl HST}/WFPC2 fields of this study (see legend at upper left).  Note the
two {\sl CXO} sources lacking optical counterparts (cross symbols).}
\end{figure}

\clearpage

\begin{deluxetable}{lllllll}
\tablecaption{Observations Catalog\label{obsvtab}}
\tablewidth{0pt}
\tablehead{
\colhead{ID}
& \colhead{$\alpha_{\rm J2000}$}
& \colhead{$\delta_{\rm J2000}$}
& \colhead{UT Date}
& \colhead{Filter}
& \colhead{$N_{\rm exp}$}
& \colhead{$T_{\rm tot}$ (s)}
\\
}
\startdata
CDFS1  & 03 32 28.21 & $-27$ 45 54.19 & 2000 Jul 22,23 & F606W & 8 & 3700
\\
&&&& F814W & 8 & 5800 \\
CDFS2  & 03 32 07.01 & $-27$ 46 36.19 & 2000 Jul 23 & F606W & 8 & 3700 \\
&&&& F814W & 8 & 5800 \\
CDFS3  & 03 32 12.11 & $-27$ 43 03.19 & 2000 Jul 27 & F606W & 8 & 3700 \\
&&&& F814W & 8 & 5800 \\
\enddata
\tablecomments{
Right ascension $\alpha$ in hours, minutes, and seconds of time.
Declination
$\delta$ in degrees, minutes, and seconds of arc.
}
\end{deluxetable}

\clearpage
\begin{deluxetable}{lcrrrrllll}
\tabletypesize{\scriptsize}
\tablecaption{Properties of CDFS (300~ksec) WFPC2
Counterparts\label{photomtab}}
\tablewidth{0pt}
\tablehead{
\colhead{CDFS ID}
& \colhead{$(\Delta\alpha,\Delta\delta)$\tablenotemark{a}}
& \colhead{$V$}
& \colhead{$I$}
& \colhead{$\log F_{\rm XS}$}
& \colhead{$\log F_{\rm XH}$}
& \colhead{$HR$}
& \colhead{$z$}
& \colhead{Optical}
& \colhead{Spectral}
\\
& \colhead{(arcsec)}
&
&
& \colhead{(cgs)}
& \colhead{(cgs)}
&
&
& \colhead{Morph.}
& \colhead{Classif.}
}
\startdata
J033202.5$-$274601 & $(-0.51,+0.08)$ & 24.68 & 22.69 & $-14.43$ & $-14.03$
& $-0.41$ &  1.613 & Ellipt. & AGN--Type~1\\
J033203.7$-$274604 & $(-0.39,+0.30)$ & 21.33 & 19.23 & $-15.30$ & $-13.95$
& $+0.57$ &  0.574 & S0 & NELG\\
J033204.5$-$274644 & \nodata\tablenotemark{b} & $>\!26.8$ & $>\!25.5$ &
$-15.44$ & $<\!-15.0$ & $-1.00$ & \nodata & \nodata & \nodata \\
J033205.4$-$274644 & \nodata\tablenotemark{b} & $>\!26.8$ & $>\!25.5$ &
$-15.87$ & $-14.88$ & $+0.24$ & \nodata & \nodata & \nodata \\
J033208.0$-$274240 & $(-1.01,-0.65)$ & 22.68 & 20.80 & $-15.43$ & $-14.80$
& $-0.18$ &  0.747 & Spiral & NELG\\
J033208.1$-$274658 & $(-0.13,-0.16)$ & 25.27 & 24.24 & $-14.71$ & $-14.29$
& $-0.40$ & \nodata & Spiral in CG & \nodata \\
J033208.3$-$274153 & $(-0.26,+0.72)$ & 25.20 & 24.27 & $-15.22$ &
$<\!-15.0$ & $-1.00$ &  2.453 & Indet. & NELG\\
J033208.7$-$274735 & $(+0.00,+0.00)$ & 19.16 & 17.80 & $-13.44$ & $-13.20$
& $-0.56$ &  0.544 & Unres. & AGN--Type~1\\
J033209.5$-$274807 & $(+0.00,-0.43)$ & 20.86 & 19.62 & $-15.11$ & $-14.34$
& $-0.02$ &  2.817 & Unres. & BAL QSO\\
J033210.6$-$274309 & $(+0.00,-0.44)$ & 24.86 & 22.12 & $-14.35$ & $-14.03$
& $-0.49$ &   2.0? & Unres. & NELG\\
J033211.0$-$274343 & $(-0.02,-0.82)$ & 22.39 & 20.54 & $<\!-16.0$ &
$-14.66$ & $+1.00$ & \nodata & Spiral & BL Lac?\\
J033211.0$-$274415 & $(+0.11,+0.00)$ & 22.38 & 21.03 & $-14.28$ & $-14.00$
& $-0.53$ &  1.606 & Unres. & AGN--Type~1\\
J033211.8$-$274629 & $(+0.02,+0.41)$ & 25.93 & 24.39 & $-15.36$ & $-14.70$
& $-0.15$ & \nodata & LSB Spiral & \nodata \\
J033212.3$-$274621 & $(-0.45,-0.14)$ & 23.74 & 21.62 & $<\!-16.0$ &
$-14.63$ & $+1.00$ &  1.033 & Spiral & \nodata \\
J033213.3$-$274241 & $(+0.29,+0.06)$ & 20.73 & 19.07 & $-15.01$ & $-14.24$
& $-0.02$ &  0.606 & Ellipt., merger? & NELG\\
J033214.0$-$274249 & $(+0.03,-0.22)$ & 22.31 & 20.43 & $<\!-16.0$ &
$-14.57$ & $+1.00$ &  0.735 & Spiral & NELG\\
J033215.0$-$274225 & $(-0.39,+0.84)$ & 23.52 & 21.72 & $-15.08$ & $-14.64$
& $-0.39$ & \nodata & Ellipt. & \nodata \\
J033216.8$-$274327 & $(-0.13,-0.49)$ & 22.85 & 21.08 & $-15.91$ &
$<\!-15.0$ & $-1.00$ & \nodata & Spiral & \nodata \\
J033217.2$-$274304 & $(+0.34,+0.26)$ & 21.54 & 19.88 & $-14.59$ & $-14.27$
& $-0.50$ & 0.561? & Ellipt. & Low S/N\\
J033226.0$-$274515 & $(+0.17,+0.00)$ & 26.44 & 25.31 & $-15.21$ &
$<\!-15.0$ & $-1.00$ & \nodata & Indet. & \nodata \\
J033228.8$-$274621 & $(+0.37,-0.28)$ & 22.93 & 20.72 & $-15.61$ &
$<\!-15.0$ & $-1.00$ &  0.750 & Ellipt. & NELG\\
J033230.1$-$274524 & $(-0.14,+0.09)$ & 22.99 & 21.35 & $-14.88$ & $-14.78$
& $-0.66$ &  2.569 & Unres. & AGN--Type~1.5?\\
J033230.1$-$274530 & $(-0.04,-0.09)$ & 21.33 & 20.43 & $-14.30$ & $-13.96$
& $-0.48$ &  1.218 & Unres. & AGN--Type~1\\
J033230.3$-$274505 & $(+0.00,+0.20)$ & 22.18 & 20.86 & $-14.38$ & $-14.06$
& $-0.49$ &  0.291 & Unres. & NELG\\
J033233.1$-$274548 & $(+0.00,-0.19)$ & 22.49 & 20.70 & $-14.91$ & $-14.52$
& $-0.43$ & 0.362?\tablenotemark{c} & Spiral in CG & \nodata \\
J033233.9$-$274521 & $(+0.46,+0.16)$ & 25.42 & 24.20 & $-15.75$ & $-14.86$
& $+0.11$ & \nodata & LSB Irr. & \nodata \\
\enddata
\tablenotetext{a}{J2000 Right Ascension $\alpha$ and Declination $\delta$
offsets ({\sl HST}$-${\sl CXO}), in arcsec.}
\tablenotetext{b}{No counterpart within 2\arcsec\ of {\sl CXO} position;
$V$, $I$ mags are $3\sigma$
upper limits.}
\tablenotetext{c}{Redshift of adjacent bright elliptical in compact group.}
\end{deluxetable}

\end{document}